# New Analytical Approach Based on Transfer Matrix Method (TMM) for Study of Tunable Plasmonic Modes in Graphene-Based Heterostructures

Mohammad Bagher Heydari [1,*], Majid Karimipour [2], Morteza Mohammadi Shirkolaei [3]

[1,*] School of Electrical Engineering, Iran University of Science and Technology (IUST), Tehran, Iran
[2] Department of Electrical Engineering, Arak University of Technology, Arak, Iran
[3] Department of Electrical Engineering, Shahid Sattari Aeronautical University of Science and Technology, Tehran, Iran

[*]Corresponding author: mo_heydari@alumni.iust.ac.ir

**Abstract:** This paper aims to study the reflection characteristics of optical beams in a hybrid graphene-hexagonal Boron Nitride (hBN)-graphene structure, which has been located on $SiO_2$-Si layers. An analytical model is presented to derive the reflection characteristics by using the transfer matrix method. The upper Reststrahlen band has been chosen as the studied frequency range. It is shown that the characteristics of the reflected beam can be effectively controlled by varying the chemical potential of graphene sheets. The obtained results represent a high value of the reflected group delay ($\tau_r = 15.3\ ps$) at the frequency of 24.9 THz. The presented investigation will be helpful to control the group delays of reflected beams and can be utilized for the design of innovative graphene-hBN devices in the mid-infrared wavelengths.

## 1. Introduction

Nowadays, graphene has attracted immense interest among scientists in nano-electronics and THz applications [1]. It has exceptional features in the mid-infrared region, such as high thermal and optical conductivities, which can be explored in many research areas of physics and chemistry. The optical conductivity of graphene opens the way to flexibly control and adjust the propagation features of plasmonic components such as couplers [2-4], filters [5-7], resonators [8-10], circulators [11-14], waveguides [15-24], sensing [25-30], and imaging [31, 32]. Graphene-based waveguides have various structures such as planar [16, 24, 33-63], cylindrical [44, 64-69], and elliptical structures [23, 70-72]. Combining graphene with other Van der Waals materials can be interesting because hybrid heterostructures have compound properties of both materials. Hexagonal Boron Nitride (hBN) is one of the famous van der Waals materials in the mid-infrared frequencies in which its permittivity function shows two kinds of phonon modes [73-75]. The hybridization of graphene with this material can generate and support new types of propagating modes called "coupled phonon-plasmon polaritons" [76-82].

Controlling the group delay of the optical beam is one of the debated topics in recent years and it has many fascinating applications such as optical buffers and delay lines [83]. In optics, some methods have been introduced and reported to achieve high levels of the group delays such as the spin Hall effect [84] and photonic crystal [85]. However, in the mid-infrared region, there is limited research related to obtaining large, tunable group delay [86]. One of the interesting ways to flexibly obtain and vary the group delay in this band is the usage of hybrid heterostructures.

Here, we propose a hybrid heterostructure composed of graphene-hBN-graphene layers. Two graphene sheets have been utilized in our system to increase the degree of freedom to adjust the reflection characteristics more flexibly. The whole structure is illuminated by a TM-polarized beam with an incident angle of θ. The analytical expressions are derived for the calculation of the reflection characteristics of the structure by using the transfer matrix method. A large value of reflected group delay, i.e. $\tau_r = 15.3\ ps$, is achievable at the frequency of 24.9 THz.

It is worthwhile to compare the proposed structure over similar configurations reported in the literature [87-92] to give a better insight into the flexibility and superiority of the presented heterostructure. In [87], the authors have studied phonon plasmon polariton modes in two Reststrahlen bands for multilayered graphene-hBN metamaterials and have reported the dispersion diagrams for these modes. A similar study is done in [88], where hyperbolic plasmon-phonon modes are examined by nano-infrared imaging. No value for reflected group delays is reported in [87, 88] because their focus is on the existence of these modes and the investigation of the propagating features. In [89], an electromagnetic absorber based on the graphene-hBN hyper crystal is proposed and authors have obtained a perfect absorber near 750 cm$^{-1}$ at the incident angle of θ=67$^0$. A new mechanism for the directed excitation of plasmon-phonon modes is suggested in [90] and negative refraction of hybrid phonon modes is presented by the same research group in [91]. In [92], the reflected group delay is reported as $\tau$=13.97 ps for the chemical potential of 0.25 eV for their structure while our obtained result is about 15.3 ps at the frequency of 24.9 THz. Therefore, our proposed structure is tunable and can change effectively the characteristics of the reflected beam by varying the chemical potential of graphene sheets.

The remainder of the paper is organized as follows. In section 2, after introducing the proposed structure, mathematical expressions will be presented for the reflected group delay. Then, in section 3, the analytical results are reported and investigated more precisely. It will be shown that the hybrid structure can flexibly control the reflected beam by changing the chemical potential of graphene sheets. Finally, section 4 concludes the article.

## 2.  The Proposed Heterostructure and its Analytical Model

Fig. 1 shows the configuration of the studied heterostructure, where the hBN layer is sandwiched between two graphene sheets and the composite structure is located on SiO$_2$-Si layers. The whole structure is illuminated by a TM-polarized beam with an incident angle of θ. There is no layer below the Si layer. The conductivity of each graphene sheet can be modeled by the following relation [93]:

$$\sigma_{1,2}\left(\omega,\mu_c,\Gamma,T\right)=\frac{-je^2}{4\pi h}Ln\left[\frac{2\left|\mu_{c,1,2}\right|-(\omega-j2\Gamma)h}{2\left|\mu_{c,1,2}\right|+(\omega-j2\Gamma)h}\right]+\frac{-je^2K_BT}{\pi h^2(\omega-j2\Gamma)}\left[\frac{\mu_{c,1,2}}{K_BT}+2Ln\left(1+e^{-\mu_{c,1,2}/K_BT}\right)\right] \quad (1)$$

In (1), $\Gamma$ is the scattering rate, $T$ is the temperature, and $\mu_{c,1,2}$ is the chemical potential of each graphene. Furthermore, $h$ is the reduced Planck's constant, $K_B$ is Boltzmann's constant, $\omega$ is radian frequency, and $e$ is the electron charge in this relation.

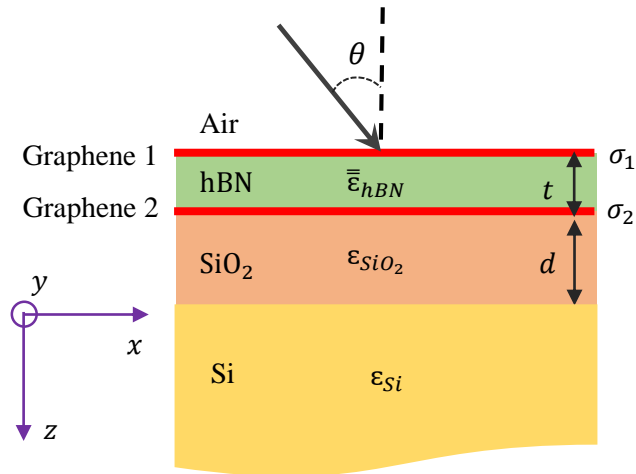

**Figure. 1.** The schematic of the studied structure.

hBN is a polar dielectric, supporting two phonon modes related to hyperbolicity, with the following permittivity tensor [74]:

$$\varepsilon_m(\omega) = \varepsilon_{\infty,m} + \varepsilon_{\infty,m} \cdot \frac{\left(\omega_{LO,m}\right)^2 - \left(\omega_{TO,m}\right)^2}{\left(\omega_{TO,m}\right)^2 - \omega^2 - j\omega\Gamma_m} \tag{2}$$

In (2), $m = \parallel$ $or$ $\perp$ is related to the transverse and z-axis, respectively. Moreover, $\omega_{LO}$, $\omega_{TO}$ show the LO and TO phonon frequencies, respectively, in which each frequency has two values in the upper and lower Reststrahlen bands: $\omega_{LO,\perp} = 24.9\ THz$, $\omega_{TO,\perp} = 23.4\ THz$, $\omega_{LO,\parallel} = 48.3\ THz$, $\omega_{TO,\parallel} = 41.1\ THz$. In (2), $\Gamma_m$ is a damping factor ($\Gamma_\perp = 0.15\ THz$, $\Gamma_\parallel = 0.12\ THz$) and $\varepsilon_m$ is related to the high-frequency permittivity ($\varepsilon_{\infty,\perp} = 4.87$, $\varepsilon_{\infty,\parallel} = 2.95$) [74]. In fig. 2, the dielectric function of hBN permittivity is depicted, which shows the lower and upper Reststrahlen bands.

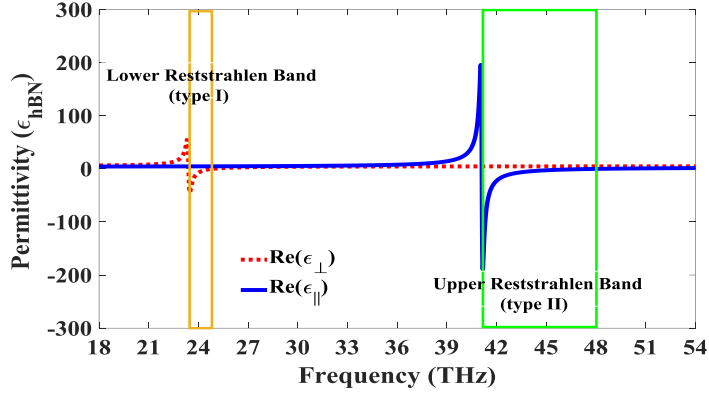

**Figure. 2.** The permittivity of hBN versus frequency. The lower and upper Reststrahlen bands are shown in this figure.

To calculate the reflection coefficient and the reflected group delay, the transfer matrix method can be utilized. For various regions, TM-polarized waves (p-polarized waves) can be written as follows:

$$\begin{aligned}
H_{y,1} &= \left(a_1 e^{ik_{z,1}z} + b_1 e^{-ik_{z,1}z}\right)e^{ik_x x} && z < 0 \\
H_{y,2} &= \left(a_2 e^{ik_{z,2}z} + b_2 e^{-ik_{z,2}z}\right)e^{ik_x x} && 0 < z < t \\
H_{y,3} &= \left(a_3 e^{ik_{z,3}z} + b_3 e^{-ik_{z,3}z}\right)e^{ik_x x} && t < z < t + d \\
H_{y,4} &= \left(a_4 e^{ik_{z,4}z} + b_4 e^{-ik_{z,4}z}\right)e^{ik_x x} && t + d < z
\end{aligned} \tag{3}$$

Thus, the transfer matrix of the whole structure is obtained:

$$\begin{bmatrix} a_1 \\ b_1 \end{bmatrix} = M \cdot \begin{bmatrix} a_4 \\ b_4 \end{bmatrix} \tag{4}$$

Where

$$M = D_{1\to2} \cdot P_2 \cdot D_{2\to3} \cdot P_3 \cdot D_{3\to4} \tag{5}$$

From Air to hBN, the transmission matrix is written as follows:

$$D_{1\to2} = \frac{1}{2}\begin{pmatrix} 1 + \eta_{TM} + \xi_{TM} & 1 - \eta_{TM} - \xi_{TM} \\ 1 - \eta_{TM} + \xi_{TM} & 1 + \eta_{TM} - \xi_{TM} \end{pmatrix} \tag{6}$$

In (6), the following parameters have been defined:

$$\eta_{TM} = \frac{\varepsilon_0 k_{2z}}{\varepsilon_\perp k_{1z}} \tag{7}$$

$$\xi_{TM} = \frac{\sigma_1 k_{2z}}{\varepsilon_0 \varepsilon_\perp \omega} \tag{8}$$

Moreover, the wave number component of the incident beam in the z-direction can be obtained by (it is supposed that the direction of the incident angle is θ):

$$k_{1z} = k_0 \cos\theta \tag{9}$$

$$k_{2z} = k_0 \sqrt{\varepsilon_\perp^2 - \frac{\varepsilon_\perp (\sin\theta)^2}{\varepsilon_\parallel}} \tag{10}$$

Now, the propagation matrix of the plasmonic wave inside the hBN layer is obtained by the following matrix (the thickness of the hBN medium is assumed to be *t*):

$$P_2 = \begin{pmatrix} e^{-jk_{2z}t} & 0 \\ 0 & e^{jk_{2z}t} \end{pmatrix} \tag{11}$$

When the beam inside the hBN medium reaches the surface of the second graphene sheet, the transmission matrix from the hBN to the SiO$_2$ layer is expressed as:

$$D_{2\to3} = \frac{1}{2}\begin{pmatrix} 1+\eta'_{TM}+\xi'_{TM} & 1-\eta'_{TM}-\xi'_{TM} \\ 1-\eta'_{TM}+\xi'_{TM} & 1+\eta'_{TM}-\xi'_{TM} \end{pmatrix} \tag{12}$$

Where the following parameters have been utilized in (12):

$$\eta'_{TM} = \frac{\varepsilon_\perp k_{3z}}{\varepsilon_{SiO_2} k_{2z}} \tag{13}$$

$$\xi'_{TM} = \frac{\sigma_2 k_{3z}}{\varepsilon_0 \varepsilon_{SiO_2} \omega} \tag{14}$$

Similar to the propagation matrix of the beam inside the hBN medium, the propagation matrix inside the SiO$_2$ layer can be written as (the thickness of the SiO$_2$ layer is assumed to be *d*):

$$P_3 = \begin{pmatrix} e^{-jk_{3z}d} & 0 \\ 0 & e^{jk_{3z}d} \end{pmatrix} \tag{15}$$

Finally, as the beam reaches the border of SiO$_2$-Si layers, the transmission matrix can be expressed as:

$$D_{3\to4} = \frac{1}{2}\begin{pmatrix} 1+\eta''_{TM} & 1-\eta''_{TM} \\ 1-\eta''_{TM} & 1+\eta''_{TM} \end{pmatrix} \tag{16}$$

In (16), the following parameters have been defined:

$$\eta''_{TM} = \frac{\varepsilon_{SiO_2} k_{4z}}{\varepsilon_{Si} k_{3z}} \tag{17}$$

By calculating the elements of the transfer matrix, the reflection coefficient and the reflectance are derived by:

$$r = \frac{M_{21}}{M_{11}} = |r(\omega)| \exp\left(j\varphi_r(\omega)\right) \tag{18}$$

$$R = |r(\omega)|^2 \tag{19}$$

For the proposed structure, $M_{21}$ and $M_{11}$ are calculated:

$$M_{11} = \left(1 + \eta_{TM} + \xi_{TM}\right).\left(1 + \eta'_{TM} + \xi'_{TM}\right).\left(1 + \eta''_{TM}\right).e^{-jk_{2z}t}.e^{-jk_{3z}d} +$$

$$\left(1 - \eta_{TM} - \xi_{TM}\right).\left(1 - \eta'_{TM} + \xi'_{TM}\right).\left(1 + \eta''_{TM}\right).e^{+jk_{2z}t}.e^{-jk_{3z}d} +$$

$$\left(1 + \eta_{TM} + \xi_{TM}\right).\left(1 - \eta'_{TM} - \xi'_{TM}\right).\left(1 - \eta''_{TM}\right).e^{-jk_{2z}t}.e^{+jk_{3z}d} +$$

$$\left(1 - \eta_{TM} - \xi_{TM}\right).\left(1 + \eta'_{TM} - \xi'_{TM}\right).\left(1 - \eta''_{TM}\right).e^{+jk_{2z}t}.e^{+jk_{3z}d}$$

(20)

$$M_{21} = \left(1 - \eta_{TM} + \xi_{TM}\right).\left(1 + \eta'_{TM} + \xi'_{TM}\right).\left(1 + \eta''_{TM}\right).e^{-jk_{2z}t}.e^{-jk_{3z}d} +$$

$$\left(1 + \eta_{TM} - \xi_{TM}\right).\left(1 - \eta'_{TM} + \xi'_{TM}\right).\left(1 + \eta''_{TM}\right).e^{+jk_{2z}t}.e^{-jk_{3z}d} +$$

$$\left(1 - \eta_{TM} + \xi_{TM}\right).\left(1 - \eta'_{TM} - \xi'_{TM}\right).\left(1 - \eta''_{TM}\right).e^{-jk_{2z}t}.e^{+jk_{3z}d} +$$

$$\left(1 + \eta_{TM} - \xi_{TM}\right).\left(1 + \eta'_{TM} - \xi'_{TM}\right).\left(1 - \eta''_{TM}\right).e^{+jk_{2z}t}.e^{+jk_{3z}d}$$

(21)

Here, if we suppose that the incident pulse is a Gaussian beam with the central and half-width of $\omega_0, \tau_0$, respectively:

$$E_i\left(0, t\right) = A_0 \exp\left(-t^2/2\tau_0^2\right)\exp\left(-i\omega_0 t\right)$$

(22)

It should be noted that the corresponding Fourier spectrum of (22) is:

$$E_i\left(0, \omega\right) = \frac{A_0 \tau_0}{2\sqrt{\pi}}\exp\left(-\left(\omega - \omega_0^2\right)^2 \tau_0^2/2\right)$$

(23)

and thus the group delay is not a function of time (t) because all relations are written in the Fourier space ($\omega$). Then, the reflected group delay is obtained:

$$\tau_r = \left[\frac{\partial \varphi_r\left(\omega\right)}{\partial \omega}\right]_{\omega = \omega_c}$$

(24)

In (24), $\omega_c$ is the carrier frequency. Now, our model is completed for the proposed heterostructure. In what follows, we will investigate the analytical results of the above mathematical relations.

## 3. Results and Discussions

This section reports the analytical results of the proposed structure. In these results, the chemical potential of graphene sheets is supposed to be $\mu_{c,1} = 0.2 \, ev, \mu_{c,2} = 0.3 \, ev$, respectively. The temperature is $T = 300 \, K$ and the relaxation time is assumed to be $\tau = 0.45 \, ps$. Both graphene layers have the similar thickness $\Delta_1 = \Delta_2 = \Delta = 0.33 \, nm$. The parameters of the hBN medium have been given in the previous section. Moreover, the geometrical parameters are $t = 100nm, d = 150nm$. The permittivity constant of $SiO_2$ and $Si$ layers are assumed to be 2.09 and 11.9, respectively. The incident angle is $\theta = 45^0$.

As explained before, there are two phonon modes in the hBN medium (see fig. 2) that are related to hyperbolicity: out-of-plane and in-plane phonon modes, which lead to two various Reststrahlen bands. First, let us consider the reflected group delays in three different frequency ranges: 24.75-25 THz (in the lower Reststrahlen band), 36.15-37.5 THz (in the middle band, i.e. between the lower and upper Reststrahlen bands) and 41.1-41.6 THz (in the upper Reststrahlen band). It should be emphasized that the resonance frequency at 24.855 THz in fig. 3 (a) is not related to the lower TO phonon frequency ($\omega_{TO,\perp} = 23.4 \, THz$). As observed in Fig. 3 (a), the group delay shows the metal-like behavior in the lower Reststrahlen band which can be enhanced. While the reflected group delay in other frequency ranges (fig. 3 (b), (c)) cannot be enhanced because it has negligible values which originate from the high values of hBN losses (imaginary part of hBN dielectric function). Therefore, we only focus on the first frequency window, i.e. 24.15-25 THz (the lower Reststrahlen band), in the following results. It should be noted that the reflected group delay in our structure originates from the Lorentz resonance mechanism.

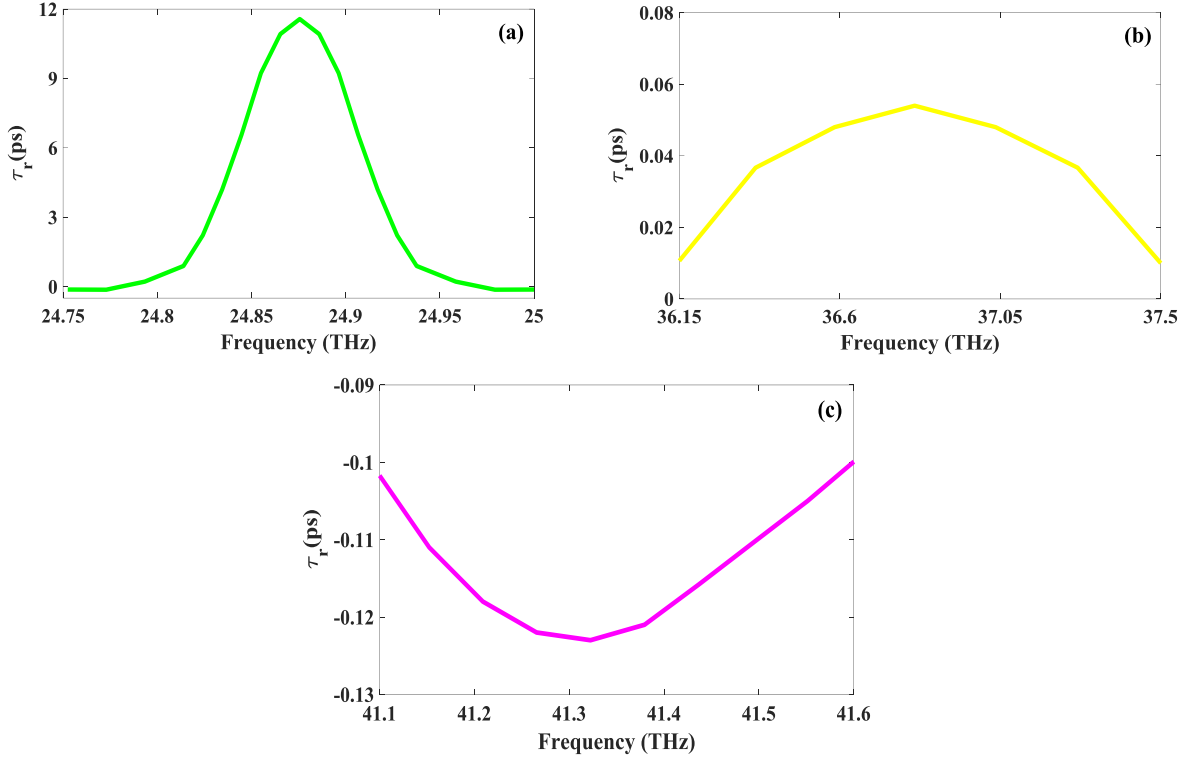

**Figure. 3.** Reflected group delay versus frequency at three frequency windows: (a) 24.75 THz-25 THz (in the lower Reststrahlen band), (b) 36.15 THz-37.5 THz (in the middle band, i.e. between the lower and upper Reststrahlen bands), (c) 41.1 THz-41.6 THz (in the upper Reststrahlen band). The chemical potential of graphene layers is supposed to be $\mu_{c,1} = 0.2\ ev$, $\mu_{c,2} = 0.3\ ev$. The thickness of the hBN and SiO$_2$ layers are 100 nm and 150 nm, respectively. The incident angle is $\theta = 45^0$.

In the previous section, we analytically obtained the elements of the transfer matrix for the proposed heterostructure. The studied structure is a tunable device in which its reflection characteristics can be varied by changing the chemical potential. Fig. 4 represents the variations of reflectance, the group delay, and the reflected phase as a function of frequency for various values of chemical potential. As noted before, the frequency range is 24.75-25 THz. The incident angle is chosen $\theta = 45^0$. It can be found from fig. 4 (a) that the reflectance has a dip of around 24.88 THz and its place can be varied as the values of chemical potential change. Around 24.88 THz, the sign of reflected phase changes, as seen in fig. 4 (b). Meanwhile, the peak of the reflected group delay varies for various values of chemical potential. A large value of reflected group delay, i.e. $\tau_r = 12.2\ ps$, is reported for the chemical potentials of $\mu_{c,1} = 0.2\ ev$, $\mu_{c,2} = 0.8\ ev$ at the frequency of 24.85 THz.

Fig. 5 shows the reflected group delay of the optical beam as a function of frequency for various values of graphene thickness. In this diagram, it is supposed that both graphene sheets have similar thicknesses ($\Delta_1 = \Delta_2 = \Delta$). The chemical potential of graphene layers are remained fixed: $\mu_{c,1} = 0.2\ ev$, $\mu_{c,2} = 0.3\ ev$. As the number of graphene layers increases (the thickness increases), the peak value of the group delay increases, as observed in fig. 5. Furthermore, one can see that the maximum peak shifts to the higher frequencies as the thickness increases. For thicker graphene sheets, a high value of group delay is achievable. For instance, the reflected group delay of 15.3ps is obtained for the thickness of $\Delta_1 = \Delta_2 = \Delta = 1\ nm$ at the frequency of 24.9 THz.

It is worth to be mentioned that the thickness of various layers in the proposed heterostructure can change the reflection characteristics of the reflected beam. One of these parameters is the thickness of the hBN layer, where its variations have been depicted in fig. 6. The chemical potential of graphene layers are $\mu_{c,1} = 0.2\ ev$, $\mu_{c,2} = 0.3\ ev$. Both graphene layers have the similar thickness $\Delta_1 = \Delta_2 = \Delta = 0.33\ nm$. Other parameters have remained fixed. It can be seen from fig. 5 that the maximum point of the reflected group delay shifts to higher frequencies as the thickness of the hBN medium increases. However, the changes are slight because the thickness of the hBN layer varies from 100nm to 120 nm.

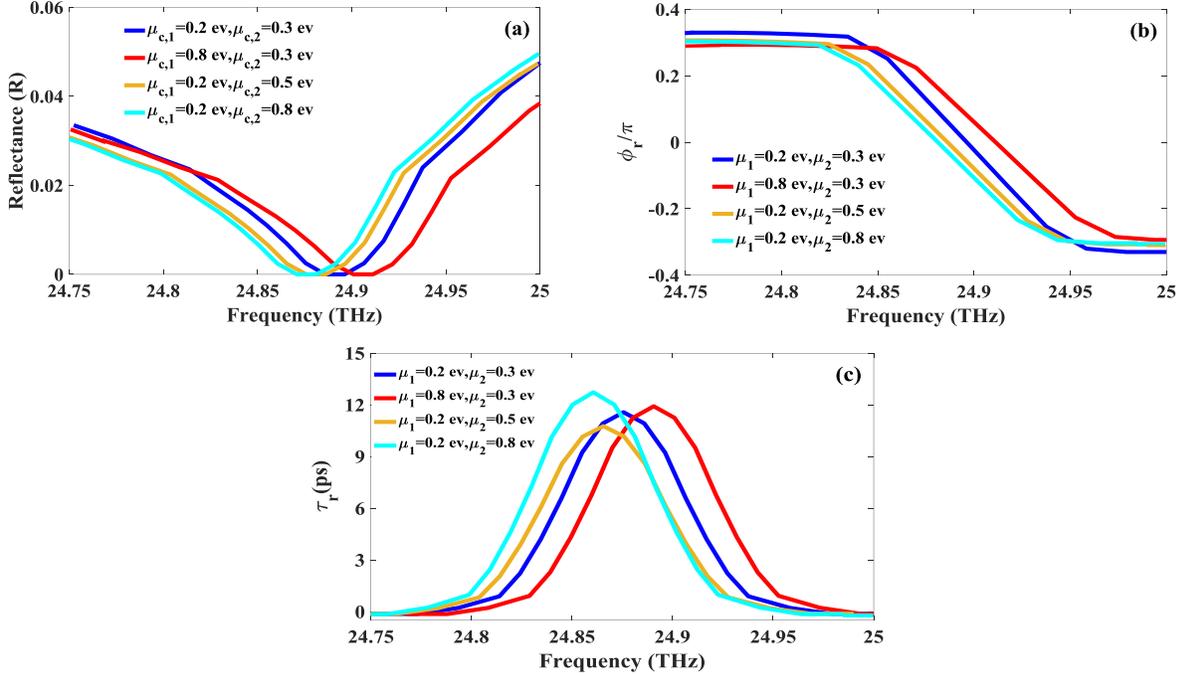

**Figure.4.** The variations of reflectance, the reflected phase, and the reflected group delay as a function of frequency in the lower Reststrahlen band, for various values of the chemical potential of graphene layers. The thickness of the hBN and SiO$_2$ layers are 100 nm and 150 nm, respectively. The incident angle is $\theta = 45^0$.

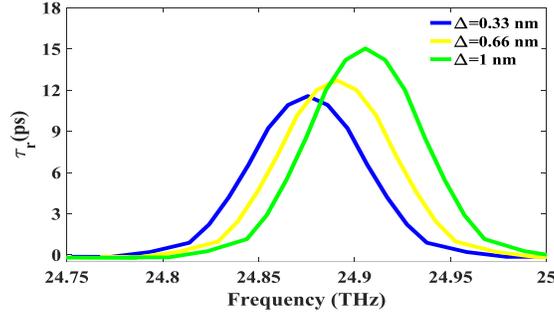

**Figure. 5.** The reflected group delay versus frequency for different values of graphene thickness. It is supposed that both graphene layers have similar thicknesses ($\Delta_1 = \Delta_2 = \Delta$). The chemical potential of graphene layers is supposed to be $\mu_{c,1} = 0.2\ ev$, $\mu_{c,2} = 0.3\ ev$. The thickness of the hBN and SiO$_2$ layers are 100 nm and 150 nm, respectively. The incident angle is $\theta = 45^0$.

As a final point, we investigate the influence of SiO$_2$ thickness on the reflected group delay. As derived in relations (16)-(21), the thickness of the SiO$_2$ layer can change the characteristics of the reflected beam. One can observe from fig. 7 that as the SiO$_2$ thickness varies from 150nm to 200 nm, the maximum peak changes, and its frequency shifts to higher frequencies. The presented study on the reflection characteristics of the reflected beam of the proposed heterostructure is useful for potential applications such as the design of optical delay lines and optical buffers.

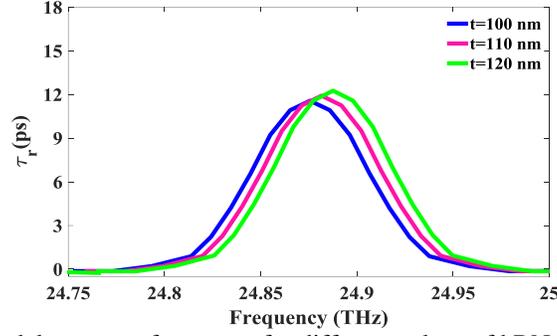

**Figure. 6.** The reflected group delay versus frequency for different values of hBN thickness. It is supposed that both graphene layers have similar thicknesses ($\Delta_1 = \Delta_2 = \Delta = 0.33\ nm$). The chemical potential of graphene layers is supposed to be $\mu_{c,1} = 0.2\ ev$, $\mu_{c,2} = 0.3\ ev$. The thickness of the SiO$_2$ layer is 150 nm. The incident angle is $\theta = 45^0$.

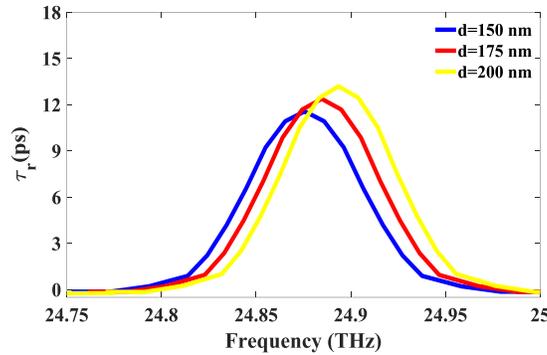

**Figure. 7.** The reflected group delay versus frequency for different values of SiO$_2$ thickness. It is supposed that both graphene layers have similar thicknesses ($\Delta_1 = \Delta_2 = \Delta = 0.33\ nm$). The chemical potential of graphene layers is supposed to be $\mu_{c,1} = 0.2\ ev$, $\mu_{c,2} = 0.3\ ev$. The thickness of the hBN layer is 100 nm. The incident angle is $\theta = 45^0$.

## 4.    Conclusion

In this article, we studied the characteristics of the reflected beam from graphene-based hBN heterostructure. Analytical expressions were obtained for calculating the reflection characteristics. A large value of the reflected group delay was seen in the lower Reststrahlen band; therefore, this frequency range was chosen to be studied. To show the tunability of the proposed structure, the variations of the reflected beam as a function of frequency were depicted and investigated for various values of chemical potential. Our results reported a large value of the reflected group delay, i.e. $\tau_r = 15.3\ ps$, at the frequency of 24.9 THz. Moreover, we showed that the thickness of graphene sheets, the hBN medium, and the SiO$_2$ layer can change the quality of the reflected beam more effectively. The authors believe that the presented study can be utilized for the design of optical delay structures in the mid-infrared region.

**Declarations**

**Ethics Approval:** Not Applicable.

**Consent to Participate:** Not Applicable.

**Consent for Publication:** Not Applicable.

**Funding:** The authors received no specific funding for this work.

**Conflicts of Interest/ Competing Interests:** The authors declare no competing interests.

**Availability of Data and Materials:** Not Applicable.

**Code availability:** Not Applicable.

**Authors' Contributions:** M. B. Heydari proposed the main idea of this work and performed the analytical modeling. M. Karimipour conducted the numerical simulations and wrote the manuscript. M. Mohammadi Shirkolaei analyzed the results and reviewed the paper.